\documentclass[pdflatex,iicol,sn-mathphys-num]{sn-jnl}

\usepackage{graphicx}%
\usepackage{multirow}%
\usepackage{amsmath,amssymb,amsfonts}%
\usepackage{amsthm}%
\usepackage{mathrsfs}%
\usepackage[title]{appendix}%
\usepackage{xcolor}%
\usepackage{textcomp}%
\usepackage{manyfoot}%
\usepackage{booktabs}%
\usepackage{algorithm}%
\usepackage{algorithmicx}%
\usepackage{algpseudocode}%
\usepackage{listings}%
\usepackage{chemformula}%
\usepackage{makecell}
\usepackage{tabularx}
\usepackage{threeparttable}

\theoremstyle{thmstyleone}%
\theoremstyle{thmstyletwo}%

\theoremstyle{thmstylethree}%

\begin{document}

\title[Paimon]{An Extensible Agentic Framework for Expert-Level Automation of Atomistic Simulations}


\author[1]{\fnm{Yutack} \sur{Park}}
\equalcont{These authors contributed equally to this work.}

\author[1]{\fnm{Yeonwoo} \sur{Chung}}
\equalcont{These authors contributed equally to this work.}

\author[1]{\fnm{Jinmu} \sur{You}}

\author[1]{\fnm{Jisu} \sur{Kim}}

\author[1]{\fnm{Suyeon} \sur{Ju}}

\author*[1,2,3]{\fnm{Seungwu} \sur{Han}}\email{hansw@snu.ac.kr}

\affil[1]{\orgdiv{Department of Materials Science and Engineering}, \orgname{Seoul National University}, \orgaddress{\city{Seoul}, \postcode{08826}, \country{Republic of Korea}}}

\affil[2]{\orgdiv{Research Institute of Advanced Materials}, \orgname{Seoul National University}, \orgaddress{\city{Seoul}, \postcode{08826}, \country{Republic of Korea}}}

\affil[3]{\orgdiv{Center for AI and Natural Sciences}, \orgname{Korea Institute for Advanced Study}, \orgaddress{\city{Seoul}, \postcode{02455}, \country{Republic of Korea}}}

\abstract{
Traditionally, atomistic simulation has been constrained by the computational scaling limits of ab initio methods and the parameterization overhead of empirical force fields. The recent emergence of universal machine learning interatomic potentials has significantly mitigated these bottlenecks, offering near-quantum accuracy and generalizability across diverse chemical spaces at a fraction of the computational cost. However, this shift has relocated the bottleneck to the human dimension: time-consuming mechanical processes, such as input preparation and data analysis, now dominate the research lifecycle. We introduce Paimon, a Platform for Agentic Integration in Materials Optimization and Nanoscale-simulations. Through hundreds of trials on an expert-level liquid electrolyte simulation, we show that Paimon substantially improves the reliability of agentic workflows by suppressing silent errors: plausible yet physically incorrect results. Paimon accommodates a new capability without modifying the agents already present, which we demonstrate by integrating an externally developed scientific agent. As an agent harness for atomistic simulations, Paimon affords researchers a continuous, science-centric workflow throughout the entire simulation lifecycle.
}

\maketitle

\section{Introduction}

Atomistic simulations are essential in materials science because they reveal the fundamental physical mechanisms and structure–property relationships that govern macroscopic behavior at the most basic level. However, they are also technically demanding and often require solid knowledge in physics and chemistry. For example, density functional theory (DFT) calculations require careful interpretation based on quantum mechanics~\cite{rev_DFT_challenge}, while classical simulations based on interatomic potentials demand specialized knowledge to accurately model large-scale systems and analyze simulation results~\cite{textbook_MD}.

Recently, the emergence of universal machine-learning interatomic potentials (uMLIPs) has begun to lower these barriers, particularly by enabling near-DFT-level accuracy at significantly lower computational cost and with reduced manual effort in potential development~\cite{rev_hsw,mlip_m3gnet,mlip_chgnet,mlip_mace-mp,mlip_sevennet,mlip_sevennet_omni}. Consequently, human effort in preparing input files, designing workflows, and analyzing data has become the primary bottleneck in the entire research pipeline in computational materials science. Automating these processes would allow less-experienced researchers to perform simulations with ease, thereby facilitating the widespread adoption of atomistic-scale modeling. In practice, end-to-end simulations are typically carried out with combinations of scientific packages that vary across domains, making an automation system that extends easily across packages valuable.

Efforts to automate atomistic simulations have gained significant momentum in recent years~\cite{auto_aflow,auto_AiiDA,auto_AMP2,auto_atomate,auto_atomate2,auto_fireworks,auto_pyiron,auto_jobflow}. These frameworks manage computation workflows, enabling their reuse across material systems and conditions by composition and parameterization of workflow nodes. Implementing new simulation scenarios, however, requires domain-specific human effort for their development~\cite{auto_semantic}. Developing reusable workflows is particularly challenging for simulations involving complex geometries or conditions, such as deposition~\cite{cutan}, where the setup is heavily system dependent.

While recent advancements in artificial intelligence (AI), in particular large language models (LLMs), have introduced the capability to generate code and inputs~\cite{cs_code_gen} for on-the-fly automation, significant challenges remain. Whereas established workflow frameworks are deterministic, an LLM probabilistically constructs the procedure, which reduces the reproducibility of the resulting workflow. The probability of failure increases substantially when LLMs are applied to multi-step, long-horizon tasks~\cite{cs_swe_bench,a_mdcrow} inherent in atomistic modeling. Furthermore, the limited parametric knowledge~\cite{rev_llm_fact,qa_mechGPT} of current LLMs often falls short of the specialized expertise required to execute sophisticated, expert-level simulations~\cite{llm_input_gen_orca,a_lammps_input_gen_sci_reports,a_genius}.

These challenges, however, are becoming surmountable with the integration of agentic AI~\cite{a_agent_survey_2025,a_review_multi_agent}. By adopting advanced workflows such as dynamic task decomposition~\cite{llm_plan_and_solve,llm_hugginggpt,a_planning,a_chemgraph} and self-reflection~\cite{llm_reflexion,llm_self_refine}, agentic AI frameworks can significantly enhance the execution accuracy of LLMs~\cite{a_agent_survey_2025,a_review_multi_agent}. Recent studies, including AtomAgents~\cite{a_pnas}, El Agente~\cite{a_ELAgente}, and CASCADE~\cite{a_cascade}, have begun to demonstrate this potential~\cite{a_aitomia,a_cascade,a_digitaldiscovery,a_dreams,a_parc,a_quasar,a_communications_materials,a_s1matagent,a_llamp,a_pnas,a_genius,a_simmof,a_chemgraph}. Nevertheless, many existing applications remain limited to relatively straightforward simulation procedures, and extensibility to new simulation domains or packages has received less attention. Furthermore, although several reports mention various errors made by agents~\cite{a_parc,a_aitomia,a_llamp}, current evaluations remain largely qualitative, with limited quantitative evaluations and systematic failure analysis.

In this work, we present Paimon (Platform for Agentic Integration in Materials Optimization and Nanoscale-simulations), a multi-agent framework designed for the end-to-end automation of atomistic simulations where computational procedures are well-developed. Paimon translates natural language queries into simulation workflows that execute tasks, analyze data, and deliver results, following the expert procedures established in the scientific literature. Central to our approach is extensibility: agents communicate through a shared interface, and simulation procedures are consolidated into natural language rather than embedded in agents. The capability of Paimon can be extended without modifying the existing agents and without predefined mappings between them. Consistent with this design, we enhance the reliability of Paimon with agentic components that are not tied to a specific simulation task or software. We quantify the contribution of each agentic component by repeated sampling and classify the failures by their detectability and severity. Finally, we demonstrate an extension of Paimon with an externally developed agent.

\begin{figure*}
\centering
  \includegraphics[width=\textwidth]{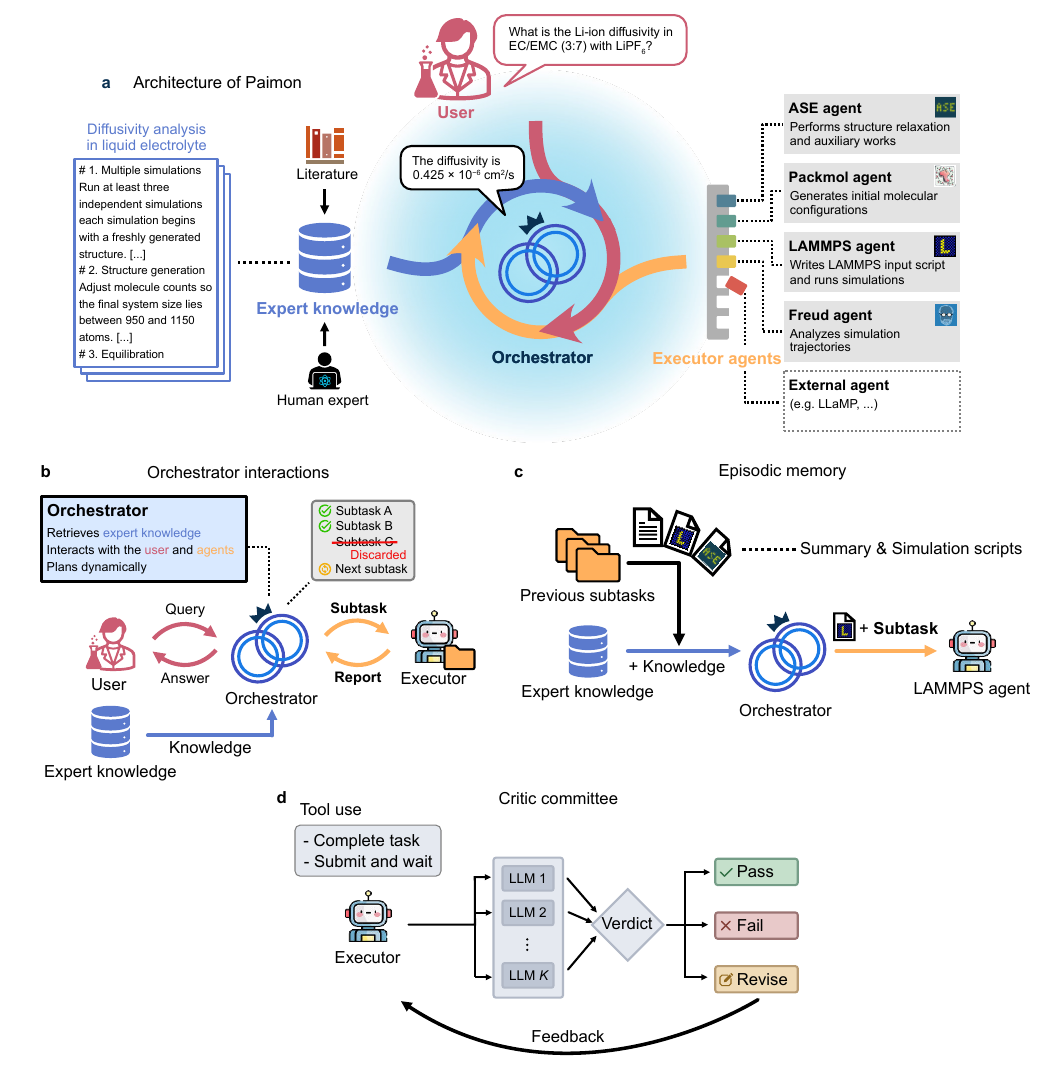}
  \caption{\textbf{Architecture of Paimon.} \textbf{a} Schematic overview of Paimon. The left panel illustrates the expert knowledge module. The right panel shows the executor agents employed in this work, from top, ASE~\cite{sw_ASE} agent, PACKMOL~\cite{sw_PACKMOL} agent, LAMMPS~\cite{sw_LAMMPS} agent, Freud~\cite{sw_freud} agent, and LLaMP~\cite{a_llamp} agent. The center panel shows a chat between the user and the orchestrator. \textbf{b} Schematic of orchestrator interactions with other modules. The orchestrator answers a user query and retrieves expert knowledge. It gives a subtask to the executor agent and receives a report from it. \textbf{c} Schematic workflow of Paimon using the episodic memory. When the orchestrator retrieves knowledge from the expert knowledge database, it is informed of a summary of a previous run and corresponding subtasks. The orchestrator then provides one of the previous subtasks as a reference example to the LAMMPS agent. \textbf{d} Critic committee module. The critic committee is invoked when the executor agent uses either the complete task or the submit and wait tool. $K$ denotes the number of independent critic LLM calls made, given the memory of the executor. The responses of the LLMs are aggregated into a verdict. The verdict is one of Pass, Fail, or Revise. For the Revise verdict, the executor receives feedback.}
  \label{fgr:arch}
\end{figure*}

\section{Results}

\subsection{Architecture of Paimon}\label{section_general}

Atomistic simulations span diverse application areas, and these domains continue to expand. Consequently, an automation system that can scale easily across these domains is essential. However, many existing agentic systems in atomistic simulations rely on hard-coded agent hierarchies and tightly scoped domain-specific tools, making their architectures difficult to extend~\cite{a_quasar}. Paimon achieves high extensibility through a centralized multi-agent architecture~\cite{a_agent_survey_2025,a_planning,a_review_multi_agent}.
Paimon consists of three independent components: expert knowledge, an orchestrator, and executors (see Fig.~\ref{fgr:arch}a). Briefly, the expert knowledge contains detailed computational procedures summarized in natural language. The orchestrator interprets the user query and controls the overall computational workflow based on this expert knowledge by managing executors. The executors are agents that handle individual scientific software packages, such as LAMMPS~\cite{sw_LAMMPS}, ASE~\cite{sw_ASE}, or PACKMOL~\cite{sw_PACKMOL}. Below, we discuss each component in more detail.

First, we begin with the orchestrator. The orchestrator serves as the central controller of Paimon by combining user queries with expert knowledge. Specifically, the orchestrator performs dynamic planning and task decomposition to manage complex simulation workflows.
The detailed role of the orchestrator in Paimon is illustrated in Fig.~\ref{fgr:arch}b. Given the user query, the orchestrator first retrieves relevant expert knowledge, a natural-language procedure guide. Specifically, it searches the knowledge database via vector retrieval~\cite{rag_vector} or extracts computational methodology if the user provides a reference research article. Once the relevant knowledge is obtained, it establishes a workflow outline and subsequently generates subtasks in an on-the-fly manner. The orchestrator can revisit earlier stages or discard subtasks depending on intermediate outcomes, allowing the system to manage complex, long-horizon research pipelines by adapting the workflow based on the specific requirements of each task.

Each subtask is assigned an executor that carries out instructions specified by the orchestrator. Upon completion of the subtask, the orchestrator receives a summary report detailing its outcome. If the orchestrator finds that the report is missing necessary information, it can consult with the executor agent for clarification. The orchestrator-executor interactions rely on a shared set of tools that are independent of executor type, allowing new executors to be added without modifying the orchestrator.

Each subtask record consists of persistent files, including the memory of the executor agent. Such records of completed subtasks constitute an episodic memory~\cite{cs_CoLAL,llm_reflexion} of Paimon, comprising a summary of the entire run and the key simulation scripts produced at each stage (see Fig.~\ref{fgr:arch}c). This memory is automatically supplied to the orchestrator alongside the expert knowledge, providing contextual awareness of prior execution history. When generating a new subtask, if the orchestrator judges that it resembles a previously completed one, it attaches the corresponding record to the executor. The attached record serves as a concrete example guiding the executor toward a previously successful history~\cite{llm_fewshot}.

Second, the executors are agents that handle specialized software packages. Atomistic simulations require highly sophisticated programming, resulting in a long-standing ecosystem of established packages specialized in structure generation~\cite{sw_ASE, sw_PACKMOL}, molecular dynamics (MD)~\cite{sw_LAMMPS}, data analysis~\cite{sw_freud,sw_MDAnalysis1,sw_MDAnalysis2}, etc. A core design philosophy of Paimon is to fully take advantage of these established scientific packages, thereby avoiding unreliable re-implementation of functionality. Accordingly, each executor agent in Paimon specializes in a specific package. For each subtask, the orchestrator issues an instruction to an executor, scoped to the capability of its package. For example, a query to the LAMMPS agent might be: ``Carry out an NVT simulation at 300 K for 1 ns and save the MD trajectory'', where NVT refers to the isothermal--isochoric ensemble.

The executor agent then autonomously manages the computational lifecycle. This includes generating input scripts or program commands, running the software or submitting jobs, validating result integrity, and postprocessing the outputs. Executors write their results to an HDF5 file~\cite{hdf5} together with the information needed to interpret them (e.g., the box convention). When an agent needs to process data produced by upstream agents (e.g., the Freud agent analyzes MD outputs of the LAMMPS agent), this HDF5 file serves as a communication channel between the agents. We refer to this mechanism as the file contract.

While the standardized format solves mechanical problems, semantic mismatches must also be resolved for different packages to interoperate. For example, particle coordinates may be wrapped into the periodic box or not, which changes how a downstream analysis treats them. Resolving such mismatches relies on the range of file formats and conventions a package supports, which must be enumerated in advance. Through the file contract, executors transfer semantic information within the data, and the downstream executor generates the necessary conversion code at runtime. Therefore, a new executor can be added without considering which agents will consume its output.

To perform these functions, each executor is equipped with dedicated tools for accessing software documentation and managing execution and I/O (see Supplementary Table~1 for tools). Among these, generating simulation scripts via retrieval-augmented generation~\cite{rag_rag,rag_review_2023} from official software documentation is particularly important. This is because the parametric knowledge of LLMs is reliable primarily for common use cases (e.g., plain NVT simulations), whereas it often fails for less common scenarios~\cite{a_aitomia,a_pnas,llm_input_gen_orca,a_genius} (e.g., deposition simulations). To further improve script quality, we use an annotator LLM that marks the most relevant sections of the retrieved document, given the subtask context.

Previous frameworks often embed simulation logic into predefined tool functions, because earlier LLMs could not reliably generate simulation scripts autonomously~\cite{a_pnas,a_mdcrow,a_lammps_input_gen_sci_reports}. The resulting systems can only perform simulations for which a corresponding tool has been implemented. The executor agents of Paimon instead generate simulation scripts via retrieval over official documentation, shifting the extensibility bottleneck from tool implementation to documentation coverage. We demonstrate in Section~\ref{sec_lammps} that this approach is now feasible for a range of tasks.

Finally, expert knowledge (see Fig.~\ref{fgr:arch}a) is structured as a collection of specialized modules that provide self-consistent, natural-language descriptions of computational procedures to guide the entire simulation process, which follows the best practices established among specialists. To respond to the user query even when it provides minimal details, expert knowledge specifies technical parameters and simulation conditions. The computational steps are organized into hierarchical sections to improve the comprehension of the overall workflow structure by the orchestrator.

While the parametric knowledge of frontier LLMs is often accurate on its own, this explicit expert knowledge base is necessary for many reasons:

i) Reliable and reproducible trajectories: Native LLM-based workflow generation is probabilistic, so the same query can yield different LLM reasoning/planning trajectories and, consequently, different computational workflows. This inconsistency in the generated approach can undermine the reliability of ensemble averages and comparisons across runs. The expert knowledge improves reproducibility by providing concrete procedures that constrain the planning path of the LLM and steer it toward consistent workflows.

ii) Enhanced accuracy: The expert knowledge modules provide precise technical parameters, such as system sizes, time steps, and MD durations, that reflect the rigor of scientific literature.

iii) Computational efficiency: There is know-how established in each application domain that enhances computational efficiency. For instance, pre-equilibration steps at high temperatures can reduce the equilibration time. By explicitly applying these techniques, one can obtain the results more quickly. 

Each expert knowledge module focuses on a specific domain but covers a broad class of materials, such as Li-ion diffusivity in liquid electrolytes or thermal conductivity in bulk materials. While the knowledge module can be authored directly by domain experts, it can also be extracted from scientific literature using LLMs, which we demonstrate in Section~\ref{section_ext}.

Additionally, Paimon introduces a critic committee (critic hereafter) module that reviews the overall agent process at each subtask (Fig.~\ref{fgr:arch}d). While prior works have incorporated similar verification layers at the planning stage~\cite{a_pnas} or for convergence checks~\cite{a_dreams}, the critic of Paimon operates at the subtask level across all executors. The critic aggregates responses from multiple LLMs through a voting mechanism to obtain a reliable verdict (see Methods for voting). For each executor, the critic gates either job submission or task completion, reviewing scripts before computational resources are consumed or validating results before a subtask is finalized (Supplementary Fig.~1). The critic returns one of three verdicts: Pass, Fail, or Revise. Upon the Revise verdict, the critic provides actionable feedback for iterative refinement~\cite{llm_self_refine}.

\subsection{Failure analysis}\label{sec_fail}

When an automated simulation system such as Paimon fails to deliver accurate results, the root cause typically falls into one of three categories. First, ``LLM Misalignment'' occurs when the foundational language model hallucinates, ignores instructions, or generates incorrect code. Second, ``System Misalignment'' happens when the overarching framework itself fails, such as by lacking the proper information in the expert knowledge. Finally, even if the AI components function perfectly, ``Fundamental and External Limitations'' can cause failures due to inaccurate data in the original reference literature, the inherent boundaries of the force fields being used, or natural numerical instabilities that crash the simulation. In practice, the first two categories constitute the primary sources of errors. We note that the boundary between LLM misalignment and system misalignment is often blurred; for instance, inherent LLM hallucinations can frequently be mitigated by providing relevant domain knowledge within its context.

The above misalignments introduce various errors, which we categorize into ``benign'', ``loud'', and ``silent'' errors depending on their ultimate impact and detectability. A benign error represents a minor procedural deviation that does not significantly compromise the physical validity of the results, such as an agent setting a simulation time step to 1~fs instead of the requested 2~fs. Conversely, remaining errors fundamentally disrupt the simulation pipeline and manifest in two distinct forms. A loud error is an explicit failure, such as configuring an extreme temperature like 3000~K (instead of the requested 300~K) that yields obviously unphysical results or specifying an invalid command option that causes the MD simulation to crash. In contrast, a silent error is more deceptive. In this case, the simulation completes without runtime errors and produces seemingly plausible data, but the underlying physical meaning is corrupted, such as misusing an analysis option that silently invalidates the outcomes.

We note that these severity boundaries form a continuous spectrum depending on the magnitude of deviation. For example, setting a molecular dynamics time step to 1~fs instead of a requested 2~fs is a benign error, reducing efficiency but preserving validity. However, using 3~fs becomes a silent error, corrupting thermodynamics without crashing. On the other hand, an extreme 10~fs becomes a loud error, immediately crashing the simulation. The observed error cases are more complex than this example, as discussed in the following sections. Silent errors are the most concerning, as they leave no signal in outputs and can only be caught by inspecting the simulation script itself.

Finally, the stochastic nature of LLMs introduces run-to-run variance~\cite{repeated_sampling}: identical prompts processed by Paimon can yield outcomes ranging from flawless execution to silent failures. Consequently, validating the reliability of LLM-driven scientific automation necessitates robust statistical analysis through repeated sampling, which in turn yields insights for enhancing system reliability.

We analyze the reliability of the major executor agent, the LAMMPS agent, and the whole system. We apply both the error severity classification and repeated sampling for their evaluation. Error classification is performed by human experts, who assign each error to one of the three severity levels. Repeated sampling is carried out by an automated pipeline in which an LLM checks each simulation script against a predefined set of binary pass or fail criteria (e.g., ``Is the time step set to 1~fs?''), which measures how often the intended simulation procedure is correctly executed across $N$ independent trials with the same prompt.

\subsection{LAMMPS agent evaluation}\label{sec_lammps}

\begin{figure*}
\centering
  \includegraphics[width=\textwidth]{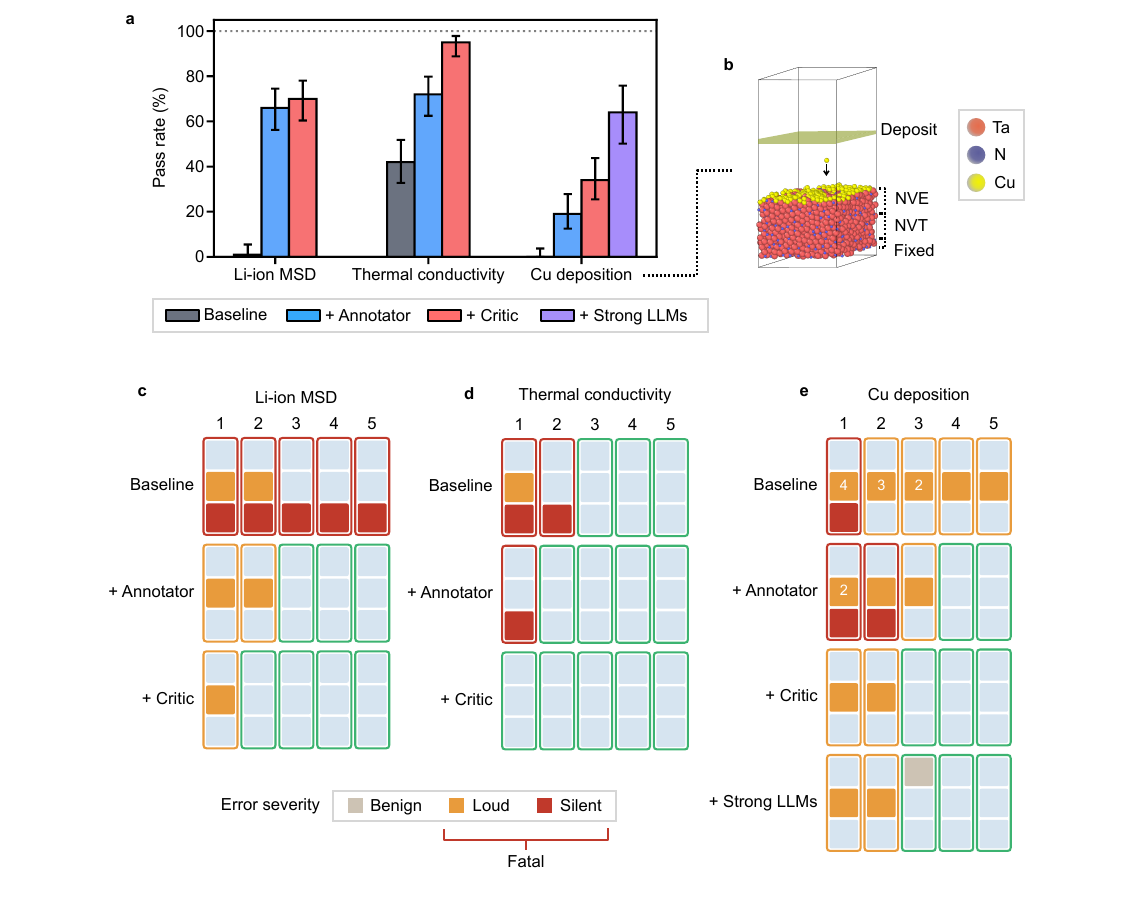}
  \caption{\textbf{LAMMPS agent evaluation.} \textbf{a} Pass rates of LAMMPS scripts generated by the LAMMPS agent for the Li-ion MSD in liquid electrolytes task, the thermal conductivity via Green--Kubo~\cite{green,kubo1957} task, and the Cu deposition on an a-Ta$_2$N slab~\cite{cutan} task. Blue, red, and violet legends denote agent configurations, each added to the previous. The strong LLMs configuration samples 50 LAMMPS scripts, while others sample 100 scripts. Error bars indicate 95\% Wilson confidence intervals~\cite{wilson_ci, wilson_ci2}. \textbf{b} Schematic diagram of the simulation cell in the Cu deposition task. NVT and NVE refer to the isothermal--isochoric and microcanonical ensembles. The fixed region contains immobilized atoms. \textbf{c--e} Human-graded subset with error severity. Five randomly sampled runs are labeled by human experts, for \textbf{c} the Li-ion MSD, \textbf{d} the thermal conductivity, and \textbf{e} Cu deposition tasks. Red, orange, and green boundaries indicate runs with silent errors, loud errors, and no or only benign errors, respectively. The overlaid numbers mark cases with more than one distinct error source. Reasoning tokens are not retained in the strong LLMs configuration. Task instructions, exemplar scripts, and rubrics are provided in Supplementary Figs.~2--7 and Supplementary Tables~2--4. Pass rates breakdown and inference costs are shown in Supplementary Figs.~8 and~9.}
  \label{fgr:lammps}
\end{figure*}

LAMMPS is a widely used molecular dynamics simulator, and many uMLIPs~\cite{mlip_m3gnet,mlip_chgnet,mlip_sevennet,mlip_mace-mp,mlip_uma} are implemented in LAMMPS. We develop a LAMMPS agent and evaluate its performance on three simulation tasks: (i) computing the mean-squared displacement (MSD) of Li-ions in a liquid electrolyte, (ii) recording the heat flux autocorrelation for the Green--Kubo~\cite{green,kubo1957}  method, and (iii) Cu deposition simulation on an a-Ta$_2$N slab~\cite{cutan}. For each simulation task, we develop simulation instructions of increasing complexity. Specifically, the Cu deposition task adapted from ref.~\cite{cutan} is the most complex, as it requires the agent to spatially partition the slab and assign different ensembles (see Fig.~\ref{fgr:lammps}b).

 We conduct ablation studies on the agent configurations by cumulatively adding the annotator and the critic to the baseline. The baseline agent uses the retrieval~\cite{rag_rag} tool for the LAMMPS document. We allocate a higher-capacity model (GPT-5~\cite{llm_gpt5}) to the annotator and critic than to the executor (GPT-5-mini), because the executor cannot correct erroneous guidance from these components. The Cu deposition task is evaluated with an extra strong LLMs configuration, which uses GPT-5.4~\cite{llm_gpt5.4} for all LLMs involved. The annotator is validated on an independent spall simulation task~\cite{annot_ood} that was not used in its development, and its guidance is correct in 187 out of 190 samples (Supplementary Table~5). In all configurations, the agent utilizes the deterministic check provided in LAMMPS to detect syntax errors before simulation.

For repeated sampling, we assess LAMMPS scripts generated by the agent using an LLM that checks each script against predefined rubrics. Independently of the rubric, human experts evaluate a subset of these scripts (50 in total) to detect failure cases that the rubric may miss and assign error severity (see Methods for details). Counting human-detected silent or loud errors as failures, the human evaluation and the repeated sampling agree on all 50 scripts, with detailed verdicts reported in Supplementary Note~1.

In all configurations, the agent submits syntactically correct LAMMPS scripts, so the remaining errors are either silent or run-time loud errors. Across all three tasks, the addition of the annotator and critic consistently improves the pass rates (Fig.~\ref{fgr:lammps}a) while suppressing both loud and silent errors (Fig.~\ref{fgr:lammps}c--e).

For the Li-ion MSD task, the pass rate significantly increases as we add the annotator, from 1\% to 66\%. This gap is attributed to the MSD analysis command. In the baseline configuration, the agent selects an incorrect option that results in a silent error. Specifically, all failed scripts in the baseline misuse the \texttt{com} option for the MSD analysis, which is incorrect because it subtracts the center of mass of the selected atoms, not the entire system. This behavior likely stems from the task instruction, which requires removing the center-of-mass linear momentum from the system. The agent conflates this condition with the \texttt{com} option. However, this failure mode becomes rare as the annotator is added.

For the thermal conductivity task, the pass rate increases from the baseline of 42\% to 72\% and 95\% as we add the annotator and then the critic, respectively (Fig.~\ref{fgr:lammps}a). The most frequent failure arises from heat flux computation, where the agent omits a command option (\texttt{virial}) required for the analysis. This case accounts for all silent errors in Fig.~\ref{fgr:lammps}d. The critic corrects this failure. The critic identifies a mismatch between the submitted LAMMPS script and the heat flux document, returning a Revise verdict citing the missing \texttt{virial} option. The agent incorporates this feedback and resubmits a revised script, which then passes the critic and subsequently the rubric.

For the Cu deposition task, the baseline does not pass, but the pass rate increases to 19\%, 34\%, and 64\% as we add the annotator, critic, and strong LLMs, respectively. The primary failure arises from the deposit region specification, where relevant guidance is split across two documents. Specifically, there is a constraint on the deposition region, which appears not in the ``region'' document but in the ``deposit'' document. Consequently, the agent assigns an incorrect deposition region that yields a loud error. This dispersal across documents limits the annotator, as it processes only a single document for each retrieval. Interestingly, the use of stronger LLMs alleviates this limitation.

Figure~\ref{fgr:lammps}c--e illustrates the error severity classified by human experts. The asymmetry of the detected errors suggests that the failures are qualitatively different. In the Li-ion MSD task with the baseline configuration, all five runs contain silent errors (Fig.~\ref{fgr:lammps}c). In contrast, the Cu deposition runs with the strong LLMs have no silent errors but two loud errors (Fig.~\ref{fgr:lammps}e). If Paimon were deployed to these tasks, the Li-ion MSD runs would return seemingly plausible but incorrect results. On the other hand, the Cu deposition runs would crash before producing complete data, while leaving the failure observable to the agent and thus recoverable.

The above analysis shows that the stochasticity of LLMs is not uniform, but it depends on the required reasoning depth. While explicit, single-step instructions yield low variance in their execution~\cite{llm_selfconsistency}, tasks demanding complex inference or the interpretation of implicit physical or technical contexts introduce significant run-to-run variance. In these scenarios, the same prompt can yield outcomes ranging from complete success to silent failures, as characterized in Fig.~\ref{fgr:lammps}c--e.

Furthermore, the necessity of the annotator highlights a mismatch between human-centric documentation and LLM processing. Because human reading speed is limited, these documents are condensed and rely on implicit domain knowledge of the reader. While LLMs can process text at far greater speeds, they often struggle to infer these implicit, cross-referenced nuances. The annotator effectively addresses this mismatch by acting as an interpretative bridge. It translates the implicit logic of human-centric manuals into the explicit directives required by the agent, thereby significantly reducing execution errors.

\subsection{End-to-end evaluation: Li-ion diffusivity in liquid electrolytes}\label{section_e2e_gated}

\begin{figure*}
\centering
  \includegraphics[width=\textwidth]{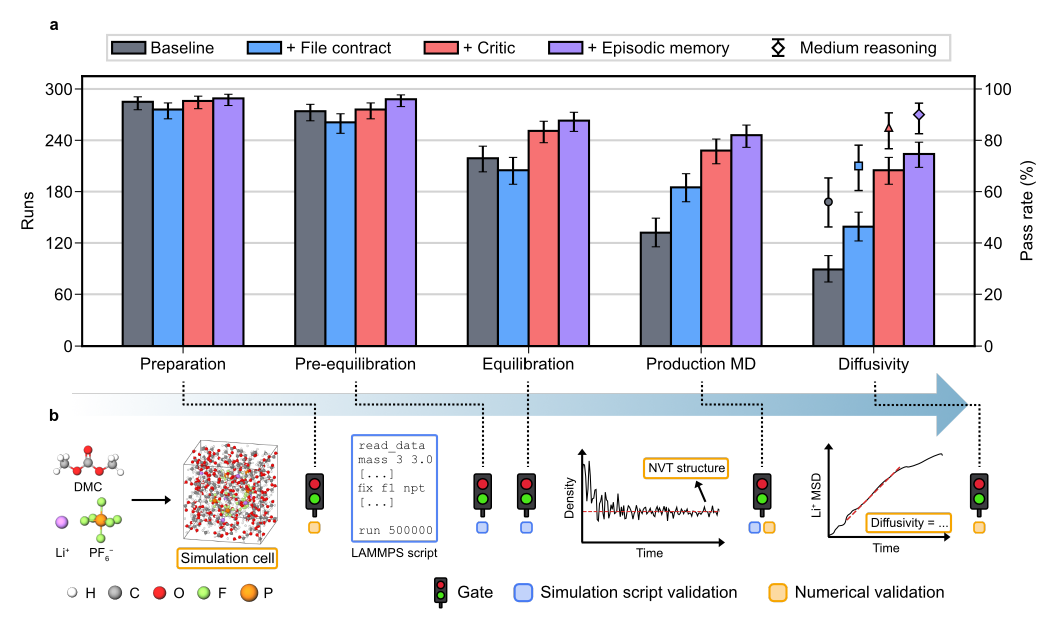}
  \caption{\textbf{Gated evaluation of the Li-ion diffusivity in liquid electrolytes task.} 
 \textbf{a} Pass rates at each stage under four cumulative configurations: baseline, file contract, critic, and episodic memory. Reasoning effort is set to ``low'' for all LLMs except the orchestrator. Each configuration is evaluated on 300 runs. Markers show pass rates of ``medium'' reasoning effort on 100 runs, with counts scaled to 300. Error bars indicate 95\% Wilson confidence intervals~\cite{wilson_ci,wilson_ci2}. \textbf{b} Schematic of Li-ion diffusivity workflow and validation gates. Each stage includes a gate with specific criteria. Blue rectangles indicate validation of the LAMMPS script by an LLM against rubrics, whereas orange rectangles indicate numerical validation of intermediate results. Gate rubrics and exemplar scripts are listed in Supplementary Tables~6--8 and Supplementary Figs.~10--12. Full results, inference costs, and failure rates by criterion are provided in Supplementary Figs.~13,~14, and Supplementary Tables~9--13, respectively.
 }
  \label{fgr:e2e}
\end{figure*}

Having established performance at the executor level, we next evaluate Paimon at the system level, which involves the orchestrator (GPT-5), inter-agent cooperation (file contract), and expert knowledge. The target task is simulating a liquid electrolyte to compute Li-ion diffusivity, strictly following the knowledge module.

We construct the knowledge module of Li-ion diffusivity using ref.~\cite{liqlyte} as a primary source. The user query to Paimon asks for Li-ion diffusivity with SevenNet-0~\cite{mlip_sevennet} alongside a solvent molecule structure. The solvent is one of diethyl carbonate (DEC), dimethyl carbonate (DMC), or propylene carbonate (PC), which are chosen for comparison with the reference.

As illustrated in Fig.~\ref{fgr:e2e}b, we decompose the simulation workflow described in the knowledge module into five stages (see Methods for computational details).
\begin{itemize}
     \item \textbf{Preparation}: The agent reconciles the molality (user query) with the atom count range (expert knowledge) to determine the molecular composition.

     \item \textbf{Pre-equilibration}: The agent substitutes the hydrogen mass with that of tritium to accelerate equilibration and performs an isothermal--isobaric (NPT) simulation.

     \item \textbf{Equilibration}: The agent restores the hydrogen mass to its physical value and resumes the NPT simulation.

     \item \textbf{Production MD}: To switch from NPT to NVT, the agent analyzes the last 0.2~ns of the NPT trajectory. It uses the frame whose density is closest to the time-averaged value as the starting point for the NVT simulation.
     
     \item \textbf{Diffusivity}: The agent analyzes the NVT trajectory to compute the Li-ion diffusivity.
 \end{itemize}

This simulation takes roughly twelve GPU hours, making repeated sampling impractical. In this regard, we introduce gated evaluation to enable an ablation study at scale. Each stage has a corresponding validation gate that Paimon must satisfy to advance. Once Paimon passes a gate, it receives the precomputed output, bypassing the expensive MD simulation. Fig.~\ref{fgr:e2e}b summarizes the validation gates for each stage, involving script and numerical validation. The script validation agrees with human grading on all 84 scripts (Supplementary Fig.~15). The full simulation result is discussed later in this section.

Starting from the baseline, we evaluate four agentic configurations by successively adding file contract, critic, and episodic memory. For each configuration, we also test ``low'' and ``medium'' reasoning effort, which controls the extent of internal reasoning in the underlying LLMs~\cite{test_time_scaling}.

Figure~\ref{fgr:e2e}a shows the gated evaluation results. At low effort, the success rate increases from 29.7\% in the baseline to 74.7\% with episodic memory, and medium effort further raises this to 90.0\%. No notable difference across solvents is observed (Supplementary Fig.~16). The progressive improvement for each addition of an agentic component suggests that they are complementary and address distinct failure modes. In the following, we examine each configuration. To isolate the contribution of each configuration to each stage, we report conditional pass rates, defined as the fraction of runs that pass a given stage among those that reached it ($n^\mathrm{pass}/n^\mathrm{reach}$).

The effect of the file contract is concentrated in the production MD stage, where the conditional pass rate increases by 29.9\% with low effort (Supplementary Fig.~13). The most frequently failed criterion is selecting an equilibrium frame for production MD. We find that the Freud agent, which performs such analysis, commits silent errors that arise from speculative parsing of simulation data. For example, the Freud agent parses the upstream LAMMPS log file under a wrong assumption, unknowingly selecting the wrong equilibrium frame (detailed in Supplementary Note~2). Classifying the production MD failures by source, with this case counted as a handoff error, we find that handoff errors disappear with the file contract at both low and medium effort (Supplementary Fig.~17).

In the diffusivity stage, the critic operates with the Freud agent rather than the previously tested LAMMPS agent. Notably, the critic increases the conditional pass rate of the diffusivity stage by 14.8\% at low effort (Supplementary Fig.~13a). This gain accompanies a reduction in silent errors, in which the computed diffusivity is wrong but numerically close to the reference. Specifically, among runs reaching the diffusivity gate, the fraction that fail while reporting a diffusivity within a factor of two of the reference decreases from 17.8\% to 7.5\%.

The episodic memory utilizes previously successful runs (see Fig.~\ref{fgr:arch}c), which we draw from different solvents than the one under evaluation. The addition of the memory raises the success rates by 6.4\% at low effort (Fig.~\ref{fgr:e2e}a). Unlike the file contract and the critic, whose gains concentrate at one or two stages, episodic memory yields modest increases at every stage. This observation aligns with its role of providing a complete prior trace as an in-context demonstration.

The success rates at medium effort all surpass their low-effort counterparts (see diffusivity of Fig.~\ref{fgr:e2e}a). Both architectural components and reasoning effort improve success rates, raising the question of whether a stronger LLM would make architectural components redundant. The monotonic decrease of the pass rate along workflow stages suggests otherwise. This decay reflects error compounding, where the success rate scales as the product of conditional pass rates. Consequently, reliability at a given workflow length does not transfer to longer ones. We therefore expect both architectural design and LLM improvements to remain essential, each extending the task length over which an agentic system operates reliably.

\begin{figure}
\centering
  \includegraphics[width=220pt]{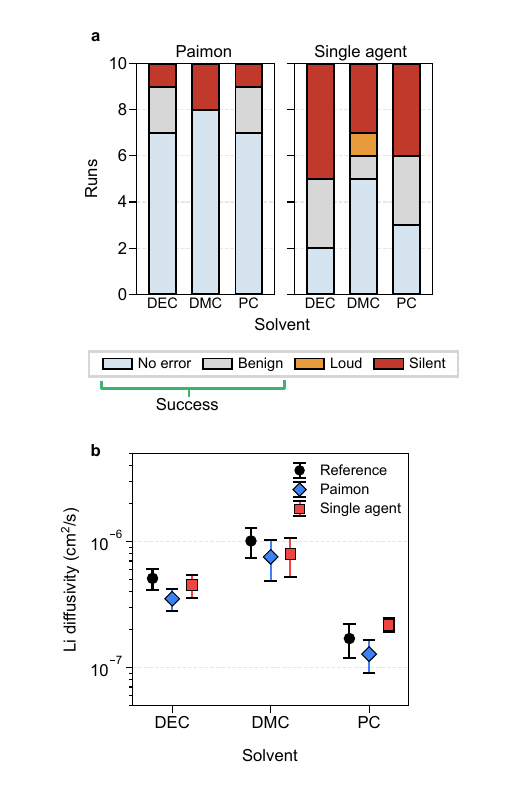}
  \caption{\textbf{Li-ion diffusivities reported by Paimon and single-agent configuration.} \textbf{a} Error severity across 30 independent runs (10 per solvent). Error severity is assigned per run based on the potential effect of the delivered result on the user. \textbf{b} Reported Li-ion diffusivities of DEC, DMC, and PC. Runs free of loud and silent errors are aggregated. The reference values are taken from ref.~\cite{liqlyte}. Error bars indicate standard deviations (the reference uses five samples). Single-agent runs for DEC and PC were performed at 298 K, whereas the reference and Paimon values are at 293 K. The inference and computational costs are provided in Supplementary Fig.~18.}
  \label{fgr:e2e_real}
\end{figure}

To examine behaviors that the gated evaluation cannot exercise, such as the statistical variance of the simulation itself, we deploy Paimon on the same workflow with the MD simulations actually executed. We run Paimon 30 times with the file contract and critic, at medium effort. We assign a single error severity to each run based on the result delivered to the user.

As shown in Fig.~\ref{fgr:e2e_real}a, 22 runs complete without errors, 4 produce benign errors, and 4 produce silent errors, with no loud errors (e.g., unphysical diffusivity). The reported diffusivity values fall in the same range as the source of the expert knowledge (Fig.~\ref{fgr:e2e_real}b).

Four out of the eight erroneous runs fall within the criteria of the gated evaluation, confirming that the failure modes detected by the gates also occur in actual execution (detailed in Supplementary Note~3). The remaining errors arise from a borderline condition where the response of the orchestrator varies. According to the knowledge module, the production MD should be extended for another 1~ns if the $R^2$ of the MSD fit, which is required for diffusivity estimation, is below 0.97. In these runs, the initial MD yields an $R^2$ slightly below this criterion.

In three runs, the orchestrator reports the unconverged diffusivity but explicitly notes that the MSD fit is inadequate and recommends extending the simulation. We classify these as benign errors. In the remaining run, the orchestrator omits such a note. We classify it as a silent error because the user has no indication that the diffusivity is unconverged. These behaviors may be influenced by the user query, as the orchestrator cited the user request for a ``quick'' simulation.

In three error-free runs, the orchestrator autonomously extends the simulation and reports a converged diffusivity. This shows that the agent can act on a conditional branch described only in natural language.

The varying behavior of the orchestrator invites a methodological observation. The round-robin study of MD simulation~\cite{round_robin} reports that researchers solving the same task can produce results deviating beyond statistical uncertainty, with the systematic component attributed to procedural choices. The variance we observe follows a similar pattern at the agent level. The knowledge module specifies the criterion and the corrective action, whereas the system prompt of the orchestrator leaves the reporting behavior underspecified. Consequently, the orchestrator resolves this gap differently across runs. This variance, unlike that among researchers, is internal to the agent and can be narrowed by refining the specification. A complementary approach is allowing user intervention, which we demonstrate in the following section.

Finally, we repeat 30 runs with a single agent to examine whether these results depend on the multi-agent architecture of Paimon. The single agent (GPT-5) holds the expert knowledge, tools, and prompts of all executors in one context. Of the 30 runs, 17 pass, one produces a loud error, and 12 produce silent errors. The loud and silent errors fall within the same set observed for Paimon. For example, the agent mishandles the frame selection in the production MD stage or produces an unphysical diffusivity caused by wrong unit conversion (loud error). While the single agent uses 55\% of the GPT-5 output tokens of Paimon (Supplementary Fig.~18), it occasionally consumes more computational time by deviating from the expert knowledge. Specifically, in four runs, the single agent skips the tritium mass substitution and doubles the computational time of the pre-equilibration stage. We note that the result does not represent the best achievable single-agent performance and thus does not imply that a multi-agent architecture is generally more reliable.

\subsection{Extensibility}\label{section_ext}

\begin{figure*}
\centering
  \includegraphics[width=\textwidth]{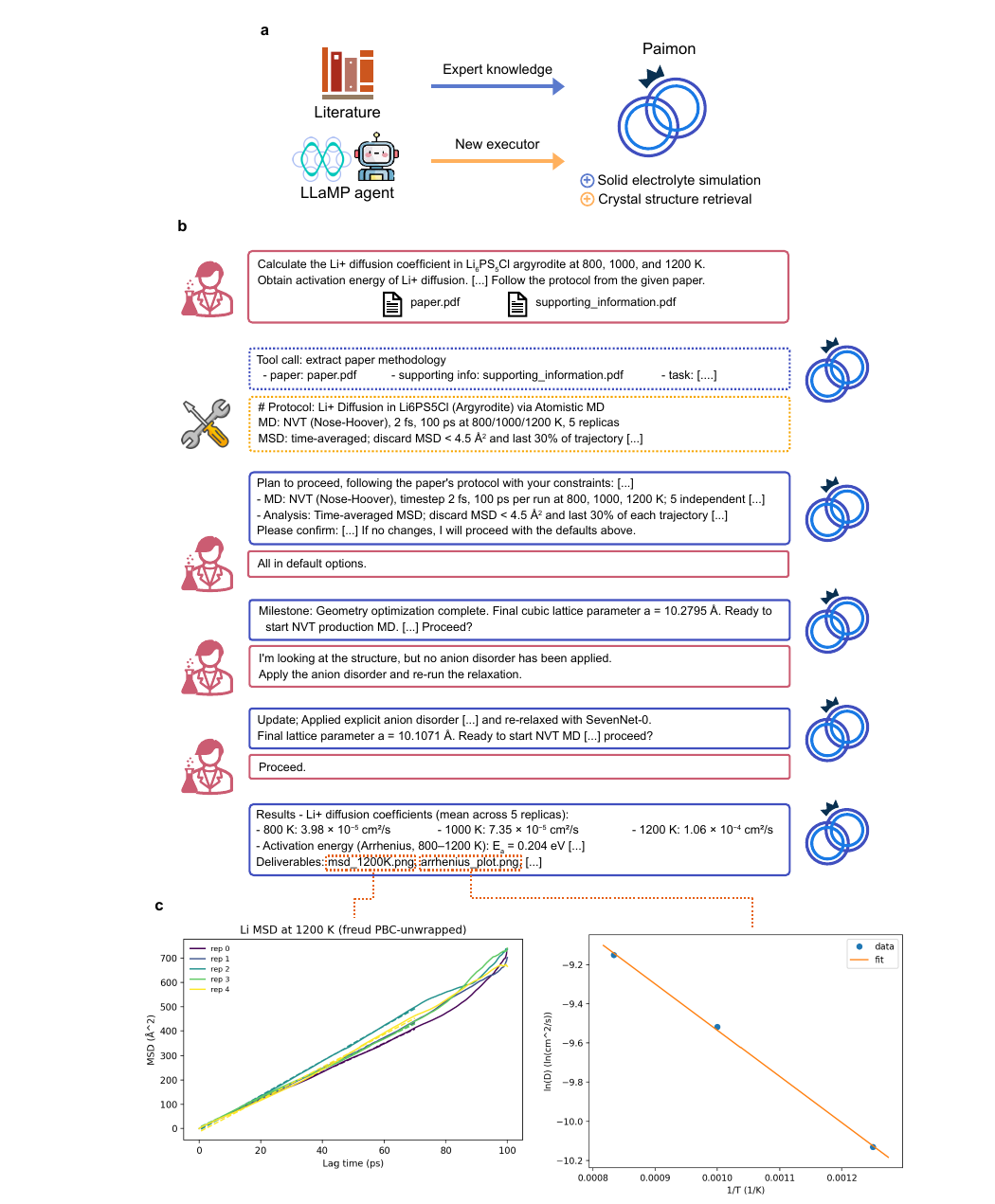}
  \caption{\textbf{Extensibility demonstrated by the Li-ion diffusivity in solid electrolyte.} \textbf{a} Schematic extension procedure. LLaMP~\cite{a_llamp} is integrated as the executor. This agent retrieves a crystalline structure from Materials Project~\cite{materials_project}. The expert knowledge module of solid electrolyte simulation is extracted from ref.~\cite{reEWC}. \textbf{b} Multi-turn interaction. Dotted boxes indicate the internal process of Paimon. Texts are lightly edited for clarity. \textbf{c} Figures generated by Paimon during the workflow. The left and right panels show the MSD plot from the MD simulation at 1200~K and the Arrhenius plot, respectively.}
  \label{fgr:solid}
\end{figure*}

Paimon achieves extensibility through two properties. First, all agents, including the orchestrator, communicate through a shared interface, so introducing a new executor requires no modification to the rest of the system. Second, the simulation procedure is consolidated into a knowledge module rather than embedded in the agents. Therefore, teaching Paimon a new simulation workflow amounts to obtaining its natural-language description.

As demonstrated in Sections~\ref{sec_lammps} and~\ref{section_e2e_gated}, run-to-run variance inherent in an agentic system makes reliable simulation execution challenging. However, this can be mitigated through multi-turn refinement, where a user provides corrective feedback to an agent~\cite{human_agent_review}.

We demonstrate the extensibility and multi-turn refinement by applying Paimon to compute the Li-ion diffusivity of \ch{Li6PS5Cl} argyrodite, aiming to reproduce the results of ref.~\cite{reEWC}. Without modifying existing agents, we integrate an externally developed LLaMP~\cite{a_llamp} agent for crystalline structure retrieval (Fig.~\ref{fgr:solid}a). The knowledge module of the argyrodite simulation is distilled from the reference by the extraction tool, which uses an LLM (GPT-5).

Figure~\ref{fgr:solid}b summarizes the result. Given the reference article, the orchestrator extracts the simulation method and follows it (see yellow box). The user points out the absence of anion disorder in the argyrodite structure, and the orchestrator incorporates this feedback. Finally, it reports the diffusivities with analysis figures (Fig.~\ref{fgr:solid}c), consistent with the reference (Supplementary Fig.~19). We note that although Paimon can execute a simulation from the literature alone, expert review through multi-turn interaction remains important when the knowledge module is newly introduced.

Because agentic systems expose natural language as their interface, they can cooperate with each other when the requirements of a task cannot be enumerated in advance. For example, Paimon can serve as a computational scientist within a general-purpose co-scientist~\cite{coscientist}, where the required calculations are derived from a research goal or hypothesis. Another natural extension is to delegate knowledge extraction to an agent specialized in scientific literature. We expect such compositions to become common as the community converges on standardized interfaces such as the Model Context Protocol~\cite{mcp_docs}. The reliability of such compositions rests on the underlying LLMs, so evaluation through repeated sampling and failure analysis remains important.

\section{Discussion}\label{section_dis}
While Paimon conveniently executes simulations from a natural-language instruction, it incurs the run-to-run variance characterized in Sections 2.3 and 2.4. Combining Paimon with workflow management systems offers deterministic methods that reduce this variance. For example, the episodic memory is currently text-based and relies on the LLMs to judge relevance. Instead, through a management system, storing workflows in a structured form would allow a calculation to be retrieved by a query rather than judged by an LLM. Similarly, on computer clusters, an agentic system would benefit from machinery such as automatic rescheduling and task resumption provided by workflow management systems~\cite{auto_AiiDA}. Then, the agent would automate the reasoning behind decisions such as parallelization parameters and node type to suit the target simulation. In general, agentic systems for atomistic simulations would benefit from deterministic mechanisms while an agentic layer is valuable where a specification cannot be conveniently written in advance.

As the capability of LLMs improves, a single-agent approach to atomistic simulation is emerging. A single agent avoids the coordination cost of a multi-agent architecture. Therefore, it is more token-efficient, as shown by El Agente Gr\'afico~\cite{a_grafico}. Furthermore, the skill mechanism of general-purpose agentic harnesses such as Claude Code makes a single agent extensible as well, since a new capability is added as a skill that the agent loads on demand, as demonstrated by AtomisticSkills~\cite{a_atom_skills}. However, the single-agent configuration evaluated here does not come with a bounded unit of work. As a result, we find that classifying errors of its runs in Section 2.4 is more laborious, as it requires reading a lengthy, continuous agent transcript. In Paimon, on the other hand, the workflow is decomposed into subtasks of bounded scope, and the critic does not conflate an issue in one stage with one in another. We expect this auditability to become more important as simulation tasks become more complex.

Finally, the end-to-end reliability of Paimon is evaluated only for the Li-ion diffusivity of liquid electrolytes, which does not establish its reliability for other systems or properties. Therefore, evaluation across a broader range of workflows is required.

\section{Methods}

\subsection{Orchestrator}

The orchestrator operates in two phases: planning and execution (Supplementary Fig.~1a). Both phases use a set of tools to manage the simulation workflow (see Supplementary Table~1 for tools). In the planning phase, the orchestrator gathers necessary knowledge and requests clarification of the user query when needed (evaluation on inappropriate user queries is reported in Supplementary Table~14). The knowledge includes an expert knowledge module and uMLIP knowledge, the latter covering available pretrained models within a specific family (e.g., SevenNet~\cite{mlip_sevennet,mlip_sevennet_omni}) and their usage details, such as the DFT functional of the training set and computational cost. Once enough information is gathered, the orchestrator calls the \texttt{outline\_plan} tool, which forwards the workflow to the execution phase.

In the execution phase, the orchestrator creates a graph that represents the simulation workflow and gains access to the subtask management tools. When creating a subtask, it specifies the instruction, dependent subtasks, an executor, and required outputs. Once a subtask is created, these inputs are consolidated into a single prompt and dispatched to the corresponding executor.
 
Upon completion of a subtask, the orchestrator receives a report. For a successful subtask, the report includes a message from the executor, the requested output values, and a summary of output structural files. This summary enumerates the composition, cell volume, and the presence of unphysically close atomic pairs (below 0.55~\AA). These quantities are computed deterministically by the framework rather than by an agent, and serve as a lightweight check on the output structure.

Unless stated otherwise, the subtask retry budget is set to one. If a subtask fails but the retry budget remains, the framework automatically retries the same subtask after resetting the executor memory. Otherwise, the orchestrator receives a failure report whose content depends on the failure mode. If the executor aborts the task on its own, the report contains the rationale of the executor; if the critic rejects the subtask, a summary of the last critic feedback is delivered. As a safeguard, the executor also terminates if it reaches a fixed iteration limit.

\subsection{Executors}

Each executor is assigned a working directory in a Linux environment with scientific packages pre-installed. Given a subtask prompt, the executor iteratively uses its tools to accomplish the task (see Supplementary Table~1 for tools). It terminates iteration by calling either the \texttt{complete\_task} or the \texttt{abort\_task} tool. Each executor maintains a short-term memory scoped to its subtask.

The LAMMPS agent generates and executes simulation scripts for LAMMPS (Large-scale Atomic/Molecular Massively Parallel Simulator)~\cite{sw_LAMMPS}. The agent retrieves a document using a hybrid retrieval scheme that combines BM25~\cite{rag_bm25} and vector retrieval~\cite{rag_vector} with reciprocal-rank fusion~\cite{rag_rrf}. Each document is indexed by the first paragraph of its Description section. A selector LLM (GPT-5-mini) chooses the final document from the top four candidates, and an annotator LLM (GPT-5) then appends guidance to the selected document, conditioned on the subtask instruction (prompt in Supplementary Note~4). When the agent uses the \texttt{write\_lammps\_script} tool, the tool automatically performs a dry run to validate the syntax.

The ASE (Atomic Simulation Environment)~\cite{sw_ASE} agent handles structure manipulation and relaxation tasks. It retrieves relevant source code via vector retrieval~\cite{rag_vector} using the code search tools adopted from CASCADE~\cite{a_cascade}. The PACKMOL~\cite{sw_PACKMOL} agent generates an atomistic structure, with the full PACKMOL documentation provided in its context. The Freud~\cite{sw_freud} agent analyzes simulation outputs, with MDAnalysis~\cite{sw_MDAnalysis1,sw_MDAnalysis2} handling trajectory I/O. For documentation access, an LLM (GPT-5-mini) reads the query together with the Freud documentation and returns guidance.

The LLaMP agent~\cite{a_llamp} is integrated for access to the Materials Project~\cite{materials_project} database. An adapter agent is implemented with a tool that sends a natural-language query to the LLaMP agent. The tool returns its response and the file it produced. The implementation consists of one Python file for the adapter, one prompt text file, and the configuration required by the LLaMP API. The prompt instructs the adapter agent to phrase queries in a form that LLaMP handles reliably. The orchestrator system prompt grows by a single line, which registers the LLaMP agent. We use the original LLaMP source code with only a minor bug fix.

The file contract is implemented through common system prompts shared across executors. Executors organize their results into \texttt{extxyz}, \texttt{dcd}, and HDF5~\cite{hdf5} formats for atomic structures, MD trajectories, and tabular data, respectively. Each HDF5 file includes links to associated output files and a description written by the executor. Downstream agents consume these files via the \texttt{inspect\_h5} tool.

\subsection{Critic committee}

The critic committee serves as a verification layer for each subtask. It operates in two modes depending on the executor: a pre-submission gate, invoked when the executor attempts to submit a job, and a completion review, invoked when the executor attempts to finish a subtask (see Supplementary Fig.~1b,c for workflow diagram). We assign the pre-submission gate to the LAMMPS agent and the completion review to the other executors.

The critic committee consists of multiple LLMs. Each LLM receives an executor agent memory, which contains the full history: the subtask instruction, tool invocations, and their outputs. The LLM is prompted to cast a vote among four labels: Pass, Concern, Reject, or Malicious. Pass indicates that the subtask instruction has been correctly followed, and no errors have been detected. Concern flags a potential issue that may compromise correctness but is not a definitive error, such as the executor failing to inspect output files before reporting. Reject identifies a definitive error that must be corrected before proceeding. Malicious indicates that the executor recognizes its task has not succeeded, yet attempts to bypass deterministic checks by fabricating results.

The critic committee aggregates the votes into a verdict. The committee returns a Fail verdict if any vote is Malicious, a Revise verdict if any vote is Reject or if the fraction of Concern votes exceeds a threshold. Otherwise, a Pass verdict is returned, and the job is submitted, or the subtask is completed and reported to the orchestrator. On a Revise verdict, an LLM summarizes the feedback from all critic LLMs that issued Reject or Concern votes. This summary is returned to the executor, which then attempts to address the identified issues within the same subtask. If a Revise verdict persists after the retry budget of the critic is exhausted, the subtask is marked as a failure.

In this work, the critic committee comprises two GPT-5 critic LLMs, and the summarizer LLM is GPT-5-mini. The critic turn budget is set to two, and the threshold for the ratio of Concern votes is set to 0.5. An evaluation of critic committee itself is reported in Supplementary Note~5.

\subsection{LAMMPS agent evaluation}\label{method_lammps_eval}
Each combination of task and configuration samples 100 LAMMPS scripts, except the strong LLMs configuration, which samples 50 scripts. 

For human evaluation, five scripts are randomly drawn from each combination, yielding 50 scripts for the error severity classification. Three human evaluators independently examine each script by executing the simulation without reference to the rubric or its result. The evaluators then discuss the detected failure cases to reach a consensus on error severity.

The LAMMPS scripts are collected at the point of job submission. In configurations where the critic is enabled, scripts are collected after passing the critic. The subtask retry budget is set to 10. GPT-5 with low reasoning effort is used to assess scripts against the rubric. 

For the thermal conductivity task, the example script in the official LAMMPS documentation is removed before building the vector database to avoid trivial retrieval. The input structure and simulation parameters for the thermal conductivity task are adapted from ref.~\cite{PbTe}. 

For the Cu deposition task, some scripts use the \texttt{jump} command. Because the control flow introduced by this command is not covered in our rubric, we manually evaluate these scripts. They account for 1, 0, 6, and 2 cases in the baseline, annotator, critic, and strong LLMs configurations, respectively.

\subsection{Li-ion diffusivity in liquid electrolyte}

The precomputed outputs and target values for the gated evaluation are prepared following the liquid electrolyte knowledge module (Supplementary Note~6 for the full module). The composition of solvents (DEC, DMC, and PC) and salts (LiPF$_6$) is determined from ref.~\cite{liqlyte}. The simulation cells are prepared using PACKMOL~\cite{sw_PACKMOL}, with a tolerance of 3~\AA\ for Li--Li pairs and 2~\AA\ for all other pairs. The initial cubic box length is determined from the summed van der Waals volumes~\cite{vdw_volume} of all atoms, $V_0$, as $1.1 \times V_0^{1/3}$. The structure is relaxed until the maximum atomic force falls below 2.0~eV/\AA.

The system is equilibrated for 1~ns in the NPT ensemble at 298~K and 1~atm using a 2~fs time step with tritium mass assigned to hydrogen atoms. The hydrogen mass is then restored, and the system is further equilibrated with a 0.4~ns NPT simulation at a 1~fs time step. The NPT trajectory is saved every 10 steps. The equilibrium density is determined by averaging the instantaneous densities over the last 0.2~ns of the NPT trajectory, and an NVT simulation is initiated from the frame whose density is closest to this value. The NVT trajectory is saved every 100 steps. During the NVT simulation, the linear momentum is zeroed every step. The Nos\'e--Hoover thermostat~\cite{nose_hoover} and barostat~\cite{npt1,npt2} are applied as implemented in LAMMPS.

The Li-ion diffusivities are computed from the MSD using the Einstein relation~\cite{liquid_textbook}, with the MSD averaged over multiple time origins~\cite{Yife_best}. The MSD fitting window is set to 10--67\% of the total simulation time, and a fit is retained if $R^2\geq 0.97$. The production trajectory length is set to 1.1~ns.

\subsection{Gated evaluation}\label{method_gated_eval}
The orchestrator receives the following user query:

``Hi, could you run a quick MD simulation at room temperature (298~K) and 1~atm for $m$ ($\pm$0.1\%) molal LiPF$_6$ in \textit{solvent}? A single full trajectory is fine, no need for replicates. I'm interested in the Li$^+$ diffusion coefficient. Use SevenNet-0 potential.''

Here, $m$ is the target molality and \textit{solvent} is one of DEC, DMC, or PC, followed by its full name in parentheses. The molality and its tolerance ($\pm$0.1\%) are chosen such that, given the range of system sizes (950--1150 atoms) specified in the expert knowledge module, the composition is uniquely determined. In evaluation without gates, the temperature is 293~K for DEC and PC while 298~K for DMC following the reference~\cite{liqlyte}.

In the baseline configuration, the Freud agent reads LAMMPS outputs directly, with the LAMMPS unit system and parsing guidance supplied in its system prompt. Such LAMMPS-specific information is replaced by HDF5 guidelines in the remaining configurations (see Supplementary Notes~7 and~8).

In the low reasoning effort setting, we conduct 100 runs for each of the three solvents, giving 300 runs in total. Episodic memory records are selected from three DEC runs of the critic configuration that passed gated evaluation. For each run with the episodic memory, one of the three records is randomly selected and supplied to the orchestrator. With these records, 150 runs are performed for DMC and PC. The medium reasoning effort setting uses 100 runs on DMC only. Its episodic memory configuration reuses the same three records sampled at low reasoning effort.

Each stage in the workflow (Fig.~\ref{fgr:e2e}) has a corresponding gate. The diffusivity gate is evaluated once the workflow completes. For the preparation gate, GPT-5-mini identifies the subtask that performs structural relaxation and locates the output structure file. GPT-5-mini performs script validation for pre-equilibration, equilibration, and production MD gates.

The pre-equilibration, equilibration, and production gates are determined by the job submission count: the first submission triggers the pre-equilibration gate, the second the equilibration gate, and the third the production MD gate. Out-of-order cases are handled as follows: a job submission before the preparation gate has been exercised counts as a preparation failure, and a fourth job submission counts as a diffusivity failure.

The preparation gate checks the relaxed structure against five criteria: (i) periodic boundary conditions enabled along all three axes, (ii) molecular composition matching the reference, (iii) maximum atomic force below 2.0~eV/\AA, (iv) minimum Li--Li distance above 2.7~\AA, (v) cubic box length within 10\% of the reference. Both the production MD and diffusivity gates perform an exact match within a numerical margin. The production MD gate passes if the input structure volume matches the reference within 0.001\%. The diffusivity gate requires the reported diffusivity to match the reference within $0.001\times10^{-6}~\mathrm{cm^2/s}$.

When the orchestrator aborts the task, the run is counted as a failure at the stage of the abort. Technical failures unrelated to task correctness, such as API errors, are discarded and rerun.

\subsection{Universal machine learning interatomic potential}\label{method_ff} 

The selection of uMLIPs is modular, and an agent retrieves potentials as needed. The choice can be specified by an expert knowledge module or by the user. If it is not specified elsewhere, the orchestrator defaults to the SevenNet~\cite{mlip_sevennet} family and autonomously selects a potential. Once the orchestrator informs the executors, they access the designated potential, such as SevenNet-Omni~\cite{mlip_sevennet_omni} or MACE-MP-0~\cite{mlip_mace-mp}, using the \texttt{switch\_venv} tool. For SevenNet potentials, executors can enable the FlashTP~\cite{sw_flashTP} kernel to accelerate simulations.

The uMLIP is specified in the user query to match the reference protocol. The liquid electrolyte task uses SevenNet-0~\cite{mlip_sevennet} with Grimme’s D3 dispersion correction with Becke--Johnson damping applied~\cite{D3,D3BJ}, following ref.~\cite{liqlyte}. The solid electrolyte task uses SevenNet-0 without the D3 dispersion correction, following ref.~\cite{reEWC}.

\subsection{Implementation}\label{method_impl}
Paimon runs on a host computer separate from the computer cluster and communicates with it over SSH. Executor working directories and files reside on the cluster. Paimon is integrated with the Slurm job scheduler~\cite{slurm}, and the standard output and error of a completed or failed job are returned to the executor that submitted it.

Each run records the transcripts of all LLMs, including the critic and the annotator. Intermediate and final output files are saved in the working directory of each executor. Software versions, including the uMLIP packages, are fixed by the Apptainer~\cite{singularity} image and do not change during a workflow.

Users interact with Paimon through a web interface, which displays the orchestrator chat and the file system of the run, from which files can be inspected and downloaded. A user preference setting controls how often and in how much detail the orchestrator requests a user response before proceeding, ranging from fully autonomous execution to confirmation of method details at each stage.

The methodology extraction is implemented with the PDF processing feature of the OpenAI API. Paimon is built on LlamaIndex. Although we use only GPT models in this work, other language models are interchangeable through LlamaIndex. Paimon retains reasoning tokens across turns to reduce token consumption, and this setting is used except in the strong LLMs configuration and in the end-to-end runs without gates.

\backmatter

\section{Data availability}
The agent transcripts, rubrics, prompts, and the data supporting all evaluations reported in this work are available at \url{https://doi.org/10.6084/m9.figshare.33972382}~\cite{data_figshare}.

\section{Code availability}
The Paimon source code is available at \url{https://github.com/MDIL-SNU/Paimon}.

\section{Acknowledgements}
The computations were carried out at the Center for Advanced Computations (CAC) at Korea Institute for Advanced Study (KIAS) and Korea Institute of Science and Technology Information (KISTI) National Supercomputing Center (KSC-2025-CRE-0284).

\section{Funding}
This work was supported by Samsung Electronics Co., Ltd. (IO250418-12669-01) and the National Research Foundation of Korea (NRF) grant funded by the Korea government (MSIT) (No. RS-2023-00247245 and No. RS-2026-25542918).

\section{Author contributions}
Y.P., Y.C., and S.H. devised and formalized the idea.
Y.P., Y.C., and J.Y. developed the code base.
S.J. authored the expert knowledge module.
Y.P., Y.C., J.Y., and J.K. contributed to the agent evaluations.
Y.P., Y.C., and S.H. prepared the manuscript.
All authors contributed to discussions and approved the paper.

\section{Competing interests}
The authors declare no competing interests.

\section{Additional information}

\bmhead{Supplementary information}
The online version contains supplementary material available at \url{doi.org}.

\bmhead{Supplementary Data 1}
The full transcript of the agent in Section~\ref{section_ext} is available as Supplementary Data 1.

\bibliography{sn-bibliography}
\end{document}